\documentstyle[ltwol,graphicx]{article}

\def\prl{{\em Phys. Rev. Lett.}\ }
\def\pra{{\em Phys. Rev.} A}
\def\prb{{\em Phys. Rev.} B}

\begin{document}

\title{QUANTUM HALL LIQUID CRYSTALS}

\author{M. M. FOGLER}

\address{Department of Physics, Massachusetts Institute of Technology,
77 Massachusetts Ave, Cambridge, MA 02139, USA\\E-mail: fogler@mit.edu}   

\twocolumn[\maketitle\abstracts{
In this report we summarize a recent progress in exploration of
correlated two-dimensional electron states in partially filled high
Landau levels. At a mean-field Hartree-Fock level they can be described
as charge-density waves, either unidirectional (stripes) or with the
symmetry of the triangular lattice (bubbles). Thermal and quantum
fluctuations have a profound effect on the stripe density wave and give
rise to novel phases, which are quite similar to smectic and nematic
liquid crystals. We discuss the effective theories for these phases,
their collective modes, and phase transitions between them.}]

\section{Charge density waves in high Landau levels}
\label{CDW}

Historically, most of the research in the area of the quantum Hall
effect has been focused on the case of very strong magnetic fields where
all the electrons reside at the lowest Landau level (LL).~\cite{FQHE} In
contrast, phenomena described below occur in moderate and weak magnetic
fields, i.e., at high LLs. Recent progress in the high LL problem can be
summarized as follows.~\cite{Fogler_xxx} The low-energy physics is
thought to be dominated by the electrons residing in the single spin
subband of the topmost ($N$th) LL, which has a filling fraction $\nu_N$
where $0 < \nu_N < 1$. All other electrons play the role of a dielectric
medium, which renormalizes the interaction among these ``active''
electrons. This picture holds at arbitrary small magnetic fields
provided there is no disorder and the temperature is zero, $T = 0$. This
is because the broadening of the $N$th LL by electron-electron
interactions is set by the quantity~\cite{Aleiner_95,Fogler_96} $E_{\rm
ex} \sim 0.1 e^2 / \kappa R_c$, where $R_c$ is the classical cyclotron
radius and $\kappa$ is the bare dielectric constant. In a metallic 2D
system with not too large $r_s$, $E_{\rm ex}$ is always smaller than the
cyclotron gap and $N$th LL is well isolated from the other LLs. The
cyclotron motion is the fastest motion in the problem, and so on the
timescale at which the ground-state correlations are established,
quasiparticles of $N$th LL behave as clouds of charge smeared along
their respective cyclotron orbits. This prompts a quasiclassical analogy
between the partially filled LL and a gas of interacting rings with
radius $R_c$ and the areal density $(N + 1 / 2) \nu_N / \pi R_c^2$. At
$\nu_N > 1 / N$ the rings overlap strongly in the real space.

\begin{figure}
\includegraphics[width=3.25in,bb=60 484 545 766]{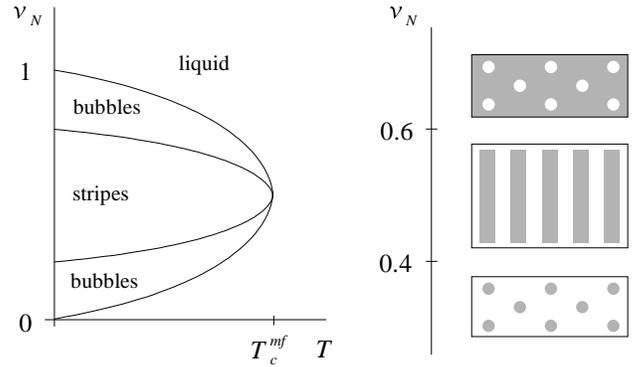}
\caption{
Left: Mean-field phase diagram.
Right: Guiding center density domain patterns at $T = 0$. Shaded and
blank areas symbolize filled and empty regions, respectively.
\label{Fig_stripes_and_bubbles}
}
\end{figure}

Within a mean-field Hartree-Fock theory a partially filled LL undergoes
a charge-density wave (CDW) transition.~\cite{Fukuyama_79} At high LLs
it occurs at a critical temperature~\cite{Fogler_96} $T_c^{m f} \sim
0.25 E_{\rm ex}$. At $0.4 < \nu_N < 0.6$ the resultant CDW is a
unidirectional, i.e., the {\it stripe phase}. At other $\nu_N$, the CDW
has a symmetry of the triangular lattice and is called the {\it bubble
phase\/}, see Fig.~\ref{Fig_stripes_and_bubbles} (left). In both cases
the CDW periodicity is set by the wavevector $q_* \approx 2.4 / R_c$. As
$T$ decreases, the amplitude of the local filling factor modulation
increases and eventually forces expulsion of regions with partial LL
occupation. The system becomes divided into depletion regions where the
local filling fraction is equal to $2 N$, and fully occupied areas where
the local filling fraction is equal to $2 N + 1$. At these low
temperatures the {\it bona fide\/} stripe and bubble domain shapes are
evident,~\cite{Fogler_96} see Fig.~\ref{Fig_stripes_and_bubbles}
(right).

%
%

The mean-field theory is expected to be valid in the quasiclassical
limit of large $N$. At moderate $N$ the CDW compete with Laughlin
liquids and other fractional quantum Hall (FQH) states. A combination of
analytic and numerical tools~\cite{Fogler_96,Fogler_97,Moessner_96,%
Rezayi_99,Shibata_01} suggests that the FQH states lose to the CDW at
$N \geq 2$.

The existence of the stripe phase as a physical reality was evidenced by
a conspicuous magnetoresistance anisotropy observed near half-integral
fractions of high LLs~\cite{Lilly_99,Du_99,Shayegan_00} (see Ref.~13 for
review). This anisotropy develops at $T < 0.1\,{\rm K}$ in high-mobility
samples. The anisotropy is the largest at total filling factor $\nu =
9/2$ ($N = 2$, $\nu_N = 1 / 2$) and decreases with increasing LL index.
At $T = 25\, {\rm mK}$ it remains discernible up to $\nu \sim 11
\frac12$ whereupon it is washed out, presumably, due to disorder and
finite temperature. The anisotropy is natural once we assume that the
stripe phase forms. The edges of the stripes can be visualized as
metallic rivers, along which the transport is ``easy.'' The charge
transfer among different edges, i.e., across the stripes, requires
quantum tunneling and is ``hard'' because the stripes are effectively
far away.

The existence of the {\it bubble phases\/} at high LLs is supported by
the discovery of reentrant integral quantum Hall effect (IQHE) at $\nu
\approx 4.25$ and $\nu \approx 4.75$. The Hall resistance at such
filling factors is quantized at the value of the nearest IQHE plateau,
while the longitudinal resistance is isotropic and shows a deep minimum
with an activated temperature dependence. The current-voltage ($I$-$V$)
characteristics exhibit pronounced nonlinearity, switching, and
hysteresis. These observations are consistent with the theoretical
picture of a bubble lattice pinned by disorder.

\section{Liquid crystal analogy for the stripe phase}

\begin{figure}
\includegraphics[width=3.1in,bb=52 357 564 760]{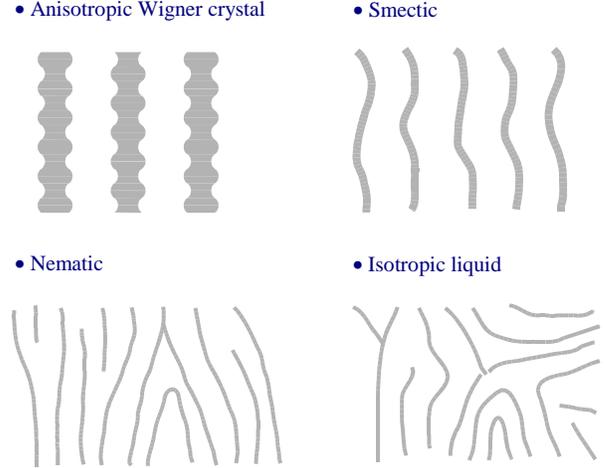}
\caption{
Sketches of possible stripe phases.
\label{Fig_four_phases}
}
\end{figure}

In the wake of the experiments, a considerable amount of work has been
devoted to the stripe phase in recent years, Refs.~14--27. It led to the
understanding that the ``stripes'' may appear in several distinct forms:
an anisotropic crystal, a smectic, a nematic, and an isotropic liquid
(Fig.~\ref{Fig_four_phases}). These phases succeed each other in the
order listed as the magnitude of either quantum or thermal fluctuations
increases. Consequently, the phase diagram of
Fig.~\ref{Fig_stripes_and_bubbles} needs modifications to incorporate
some of those phases. The general structure of the revised phase diagram
for the quantum ($T = 0$) case was discussed in the important paper of
Fradkin and Kivelson.~\cite{Fradkin_99} However, pinpointing the new
phase boundaries in terms of the conventional parameters $r_s$, $\nu$,
and $T$ requires further work. The most intriguing are the phases which
bear the liquid crystal names: the smectic and the nematic. They are the
main subject of this report. Let us start with the basic definitions of
these phases.

The smectic is a liquid with the 1D periodicity, i.e., a state where the
translational symmetry is spontaneously broken in one spatial
direction.~\cite{DeGennes_book} The rotational symmetry is of course
broken as well. An example of such a state is the original Hartree-Fock
stripe solution~\cite{Fogler_96} although a stable quantum Hall smectic
must have a certain amount of quantum fluctuations around the mean-field
state.~\cite{MacDonald_00,Fertig_99} The necessary condition for the
smectic order is the continuity of the stripes. If the stripes are
allowed to rupture, the dislocations are created. They destroy the 1D
positional order and convert the smectic into the nematic.~\cite{Toner_81}

By definition, the nematic is an anisotropic
liquid.~\cite{DeGennes_book} There is no long-range positional order. As
for the orientational order, it is long-range at $T = 0$ and
quasi-long-range (power-law correlations) at finite $T$. The nematic is
riddled with dynamic dislocations. Other types of topological defects,
the disclinations, may also be present but remain bound in pairs, much
like vortices in the 2D $X$-$Y$ model. Once they unbind, all the spatial
symmetries are restored. The resultant state is an isotropic liquid with
short-range stripe correlations. As the fluctuations due to temperature
or quantum mechanics increase further, it gradually crosses over to the
``uncorrelated liquid'' where even the local stripe order is
obliterated.

%
%

It is often the case that the low-frequency long-wavelength physics of
the system is governed by an effective theory involving a relatively
small number of dynamical variables. In the remaining sections we will
discuss such type of theories for the quantum Hall liquid crystals.

\section{Smectic state}

{\it Effective theory\/}.--- The collective variables in the smectic are
(i) the deviations $u(x, y)$ of the stripes from their equilibrium
positions and (ii) long-wavelength density fluctuations $n$ about the
average value $n_0$. The latter fluctuations may originate, e.g., from
width fluctuations of the stripes. Let us assume that the stripes are
aligned in the $\hat{\bf y}$-direction, then the symmetry considerations
fix the effective Hamiltonian for $u$ and $n$ to
be~\cite{DeGennes_book,Fogler_00}
\begin{equation}
 H = \frac{Y}{2} \Bigl[\partial_x u - \frac12 (\nabla u)^2\Bigr]^2
   + \frac{K}{2} (\partial_y^2 u)^2  + \frac12 n U n,
\label{H_smectic}
\end{equation}
where $Y$ and $K$ are the phenomenological compression and the bending
elastic moduli, and $U(r) = e^2 / \kappa r$ should be understood as the
integral operator. The dynamics of the smectic is dominated by the
Lorentz force and is governed by the Largangean
\begin{equation}
 {\cal L} =  p \partial_t u - H,\quad
  \partial_y p = -m \omega_c (n + n_0 \partial_x u),
\label{L_smectic}
\end{equation}
where $m$ is the electron mass and $\omega_c = e B / m c$ is the cyclotron
frequency.

From Eqs.~(\ref{H_smectic}) and (\ref{L_smectic}) we can derive the
spectrum of collective modes, the {\it magnetophonons\/}. It is natural
to start with the harmonic approximation where one replaces the first
term in $H$ simply by $(Y / 2) (\partial_x u)^2$. Solving the equations
of motion for $n$ and $u$ we obtain the magnetophonon dispersion
relation:~\cite{Fogler_00}
\begin{equation}
\omega({\bf q}) =  \frac{\omega_p(q)}{\omega_c} \frac{q_y}{q}
\left[\frac{Y q_x^2 + K q_y^4}{m n_0}\right]^{1/2}.
\label{omega_smectic}
\end{equation}
Here $\omega_p(q) = [n_0 U(q) q^2 / m]^{1/2}$ is the
plasma frequency and $\theta = \arctan (q_y / q_x)$ is the angle between
the propagation direction and the $\hat{\bf x}$-axis. For Coulomb
interactions $\omega_p(q) \propto \sqrt{q}$. Unless propagate nearly
parallel to the stripes, $\omega({\bf q})$ is proportional to $\sin 2
\theta\, q^{3 / 2}$. One immediate consequence of this dispersion is
that the largest velocity of propagation for the magnetophonons with a
given $q$ is achieved when $\theta = 45^\circ$.

{\it Thermal fluctuations and anharmonisms\/}.--- From
Eq.~(\ref{H_smectic}) we can readily calculate the mean-square
fluctuations of the stripe positions at finite $T$, e.g.,
\begin{equation}
\langle [u(0, 0) - u(0, y)]^2 \rangle = \frac{k_B T}{2 \sqrt{Y K}} |y|.
\label{uu_smectic}
\end{equation}
As one can see, at any finite temperature magnetophonon fluctuations are
growing without a bound; hence, the positional order of a 2D smectic is
totally destroyed~\cite{DeGennes_book} at sufficiently large distances
along the $\hat{\bf y}$-direction, $|y| \gg \Lambda \sqrt{Y K} / k_B T
\equiv \xi_y$ where $\Lambda = 2 \pi / q_*$ is the interstripe
separation. Similarly, along the $\hat{\bf x}$-direction, the positional
order is lost at lengthscales larger than $\xi_x = (Y / K)^{1/2}
\xi_y^2$.

\begin{figure}
\center\includegraphics[width=2.8in,bb=63 537 560 765]{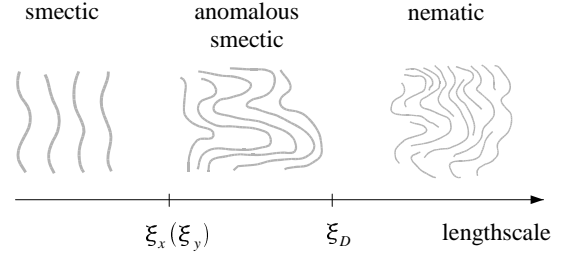}
\caption{
Portraits of the stripe phase on different lengthscales.
\label{Fig_spaghetti}
}
\end{figure}

Another type of excitations, which disorder the stripe positions are
the aforementioned dislocations. The dislocations in a 2D smectic have a
finite energy $E_D \sim K$. At $k_B T \ll E_D$ the density of thermally
excited dislocations is of the order of $\exp(-E_D / k_B T)$ and the
average distance between dislocations is $\xi_D \sim \Lambda \exp(2 k_B
T / E_D)$. At low temperatures $\xi_x, \xi_y \ll \xi_D$; therefore, the
following interesting situation emerges (Fig.~\ref{Fig_spaghetti}). On
the lengthscales smaller than $\xi_y$ (or $\xi_x$, whichever
appropriate) the system behaves like a usual smectic where
Eqs.~(\ref{H_smectic}--\ref{omega_smectic}) apply. On the lengthscales
exceeding $\xi_D$ it behaves\footnote{In a more precise
treatment,~\cite{Golubovic_92} the lengthscales $\xi_{D x} \propto
\xi_D^{6/5}$ and $\xi_{D y} \propto \xi_D^{4/5}$ are introduced such
that $\xi_{D x} \xi_{D y} = \xi_D^2$.} like a nematic.~\cite{Toner_81}
In between the system is a smectic but with very unusual properties. It
is topologically ordered (no dislocations) but possesses enormous
fluctuations. In these circumstances the harmonic elastic theory becomes
inadequate and anharmonic terms must be treated carefully.

%
%

As shown by Golubovi\'c and Wang,~\cite{Golubovic_92} the anharmonisms
cause power-law dependence of the parameters of the effective theory on
the wavevector ${\bf q}$:
\begin{equation}
 Y \sim Y_0 (\xi_y q_y)^{1/2},\quad  K \sim K_0 (\xi_y q_y)^{-1/2},
\label{limit_A}
\end{equation}
for $q_x \ll \xi_x^{-1} (q_y \xi_y)^{3/2}$, $q_y \ll \xi_y^{-1}$, and
\begin{equation}
 Y \sim Y_0 (\xi_x q_x)^{1/3},\quad  K \sim K_0 (\xi_x q_x)^{-1/3},
\label{limit_B}
\end{equation}
for $q_x \ll \xi_x^{-1}$ and $q_y \ll \xi_y^{-1} (q_x \xi_x)^{2/3}$. The
lengthscale dependence of the parameters of the effective theory is a
common feature of fluctuation-dominated phenomena. It should be
mentioned that the lower critical dimension for the smectic order is $d
= 3$,~\cite{DeGennes_book} so that the 2D smectic is {\it below\/} its
lower critical dimension. This is the reason why the scaling
behavior~(\ref{limit_A}) and (\ref{limit_B}) does not persist
indefinitely but eventually breaks down above the lengthscale $\xi_D$
where the crossover to the thermodynamic limit of the nematic behavior
commences.

The scaling shows up not only in the static properties such as $Y$ and
$K$ but also in the dynamics. The role of anharmonisms in the dynamics
of conventional 3D smectics has been investigated by Mazenko {\it et
al.\/}~\cite{Mazenko_83} and also by Kats and
Lebedev~\cite{Kats_Lebedev}. For the quantum Hall stripes the analysis
had to be done anew because here the dynamics is totally different. It
is dominated by the Lorentz force rather than a viscous relaxation in
the conventional smectics. This task was accomplished in
Ref.~17. The calculation was based on the
Martin-Siggia-Rose formalism combined with the $\epsilon$-expansion
below $d = 3$ dimensions. One set of results concerns the spectrum of
the magnetophonon modes, which becomes
\begin{equation}
\omega({\bf q}) \sim \sin \theta \cos^{7/6}\theta\,
(\xi_x q)^{5/3} \frac{\omega_p(\xi_x^{-1})}{\omega_c \xi_x}
\sqrt{\frac{Y_0}{m n_0}}.
\label{omega_m_R}
\end{equation}
Compared to the predictions of the harmonic theory,
Eq.~(\ref{omega_smectic}), the $q^{3/2}$-dispersion changes to
$q^{5/3}$. Also, the maximum propagation velocity is achieved for the
angle $\theta \approx 53^\circ$ instead of $\theta = 45^\circ$. These
modifications, which take place at long wavelengths, are mainly due to
the renormalization of $Y$ in the static limit and can be obtained
by combining Eqs.~(\ref{omega_smectic}) and (\ref{limit_B}). Less obvious
dynamical effects peculiar to the quantum Hall smectics include a novel
dynamical scaling of $Y$ and $K$ as a function of frequency and a
specific $q$-dependence of the magnetophonon damping.~\cite{Fogler_00}

The latter issue touches on an important point. Our effective theory
defined by Eqs.~(\ref{H_smectic}) and (\ref{L_smectic}) is based on the
assumption that $u$ and $n$ are the only low-energy degrees of freedom.
It is probably well justified at $T \rightarrow 0$ but becomes incorrect
at higher temperatures. The point of view taken in Ref.~17 is that in
the latter case thermally excited quasiparticles (``normal fluid'')
should appear and that they should bring dissipation into the dynamics
of the magnetophonons. Another intriguing possibility is for
quasiparticles or other additional low-energy degrees of freedom to
exist even at $T = 0$. Such more complicated smectic states are
interesting subjects for future study.

\section{Nematic}

As discussed above, at finite temperature and in the thermodynamic
limit, the smectic phase is always unstable. The lowest degree of
ordering is that of a nematic.~\cite{DeGennes_book} An intriguing
possibility~\cite{Fradkin_99} is to have a nematic phase already at $T =
0$, due to quantum fluctuations. The collective degree of freedom
associated with the nematic ordering is the angle $\phi({\bf r}, t)$
between the local normal to the stripes ${\bf N}$ and the $\hat{\bf
x}$-axis orientation. The effective Hamiltonian for ${\bf N}$ is
dictated by symmetry to be
\begin{equation}
H_N = \frac{K_1}{2} (\nabla {\bf N})^2
  + \frac{K_3}{2} |\nabla \times {\bf N}|^2.
\label{H_nematic}
\end{equation}
The phenomenological coefficients $K_1$ and $K_3$ are termed the splay
and the bend Frank constants.~\cite{DeGennes_book} A particularly simple
form is obtained if $K_1 = K_3$, in which case $H_N = (K_3 / 2) (\nabla
\phi)^2$ just like in the $X$-$Y$ model.

Another obvious degree of freedom in the nematic are the density
fluctuations $n({\bf r}, t)$. A peculiar fact is that in the static
limit $n$ is totally decoupled from ${\bf N}$, and so it does not enter
Eq.~(\ref{H_nematic}). However, since the nematic is less ordered than
even a smectic, the question about extra low-energy degrees of freedom
or additional quasiparticles is highly relevant. We believe that
different types of quantum Hall nematics are possible in nature. In the
simplest case scenario ${\bf N}$ and $n$ are the only low-energy degrees
of freedom. This is presumably the case when the nematic order is a
superstructure on top of a parent uniform state. A concrete example
is described by a wavefunction proposed by Musaelian
and Joynt:~\cite{Musaelian_96}
\begin{equation}
\Psi = \prod\limits_{j < k} (z_{j} - z_{k}) [(z_{j} - z_{k})^2 - a^2]
\times \exp \Bigl(-\sum_j |z_j|^2 / 4 l^2\Bigr).
\label{nematic_trial_wavefunction}
\end{equation}
Here $z_j = x_j + i y_j$ is a complex coordinate of $j$th electron, $l =
\sqrt{\hbar c / e B}$ is the magnetic length, and $a$ is another complex
parameter that determines the degree of orientational order and the
direction of the stripes. The rotational invariance is broken if $|a|$
exceeds some critical value. This particular wavefunction corresponds to
$\nu = \frac13$ but can be easily generalized to higher Landau levels
with $\nu_N = \frac13$ filling. This type of state has been studied by
Balents~\cite{Balents_96} and recently by the present
author.~\cite{Fogler_01} It was essentially postulated that the
effective Largangean takes the form 
\begin{equation}
{\cal L} = \frac12 \gamma^{-1} (\partial_t {\bf N})^2 - H.
\label{L_nematic}
\end{equation}
(As hinted above, the full expression contains also couplings between
$\partial_t {\bf N}$ and mass currents but they become vanishingly small
in the long-wavelength limit). The collective excitations are
charge-neutral fluctuations of the director. They have a linear
dispersion,
\begin{equation}
\omega({\bf q}) =  q \sqrt{{K_1}{\gamma} \cos^2 \theta +
                           {K_3}{\gamma} \sin^2 \theta},
\label{omega_nematic}
\end{equation}
and resemble spinwaves in the $X$-$Y$ quantum rotor model.

\begin{figure}
\center\includegraphics[width=2.2in,bb=122 265 568 513]{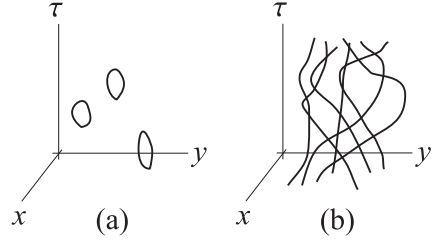}
\vspace{0.1in}
\setlength{\columnwidth}{3.2in}
\caption{
Worldlines of dislocations in (a) smectic (b) nematic.
\label{Fig_worldlines}
}
\end{figure}

The quantum nematic phase must be separated from the stable
zero-temperature smectic by a quantum phase transition.
Further insights into the properties of the quantum nematics
can be gained by analyzing the nature of such a transition.
By analogy to the classical smectic-nematic transition in
two~\cite{Toner_81} and three~\cite{Toner_82} dimensions, we expect the
quantum one to also be driven by the proliferation of dislocations.
Pictorially, the difference between the smectic and nematic can be
represented as follows. The dislocations are viewed as lines in the (2 +
1)D space. In the smectic phase, they form small closed loops
(Fig.~\ref{Fig_worldlines}a) that depict virtual pair
creation-annihilation events; in the nematic phase arbitrarily long
dislocation worldlines exist and may entangle
(Fig.~\ref{Fig_worldlines}b), similar to worldlines of particles in a
Bose superfluid.

%
%

To incorporate the dislocations into the effective
Lagrangean~(\ref{L_smectic}), we use a well known duality
transformation, see, for example, Ref.~34. By means of such a
transformation the original degrees of freedom $u$ and $n$ are traded
for new variables: the second-quantized dislocation field $\Phi$ and an
auxillary $U(1)$ gauge field $a_\mu$, which mediates the interaction
among the dislocations. The imaginary-time effective action for $\Phi$
and $a_\mu$ has the form
\begin{eqnarray}
A &=& \int\limits_0^\beta d \tau \int d^2 {\bf r}
 \biggl\{\frac{t_\mu}{2}|(-i \partial_\mu - \Lambda a_\mu
 - e_D a_\mu^{\rm ext}) \Phi|^2\biggr.
\nonumber\\
\mbox{} &+& \biggl.V(\Phi) + H_a[a]\biggr\},
\label{A_dual}\\
H_a &=& \frac{\sigma_x^2}{2 Y}
+ \sigma_y (-2 K \partial_y^2)^{-1} \sigma_y
+ \frac{l^4}{2} \partial_y \sigma_\tau U \partial_y \sigma_\tau,
\label{H_a}\\
\sigma_\mu &=& \epsilon_{\mu\nu\lambda}
\partial_\nu a_\lambda \equiv [\partial \times a]_\mu.
\label{a}
\end{eqnarray}
Phenomenological parameters introduced above are as follows. Parameter
$t_\tau$ is of the order of $\hbar^2 / E_c$, where $E_c$ is the
dislocation core energy estimated within the Hartree-Fock
approximation in Refs.~4 and 21. Unfortunately, this estimate is not
reliable in the quantum nematic state where the quantum fluctuations are
large. Parameter $t_x$ of dimension of ${\rm energy} \times {\rm
(length)}^2$ is the hopping matrix element for dislocation motion in the
$\hat{\bf x}$-direction, i.e., dislocation {\it glide\/}. Such a glide
requires quantum tunneling and is exponentially small unless $\Lambda <
l$. Parameter $t_y$ describes the dislocation climb, which also
originates from the dynamics on the microscopic length scales. Yet
another phenomenological variable is the potential $V(\Phi) = m_\Phi
|\Phi|^2 + r_\Phi |\Phi|^4 + \ldots$, which accounts for a self-energy
and a short-range interaction between the dislocations; the scales of
$m_\Phi$ and $r_\Phi$ are set by $E_c$ and $E_c \Lambda^2$,
respectively. Finally, $e_D$ is electric charge of the dislocation that
couples to the external vector potential $a_\tau^{\rm ext} = a_x^{\rm
ext} = 0$, $a_y^{\rm ext} = B x$. This coupling is introduced only for
the sake of generality. Since we study electron liquid crystal phases
derived from incompressible liquids, we expect dislocations to be
electrically neutral, i.e., $e_D = 0$.

To recover Eq.~(\ref{omega_nematic}) we assume that the dislocations
have condensed, $\langle \Phi \rangle = \Phi_0 \neq 0$. Solving for the
collective mode spectrum of the action~(\ref{A_dual}), we find
\begin{equation}
\omega_1({\bf q}) =
 \left(\frac{m_x}{m_\tau} q_x^2 + m_x K q_y^2\right)^{1/2},
\:\: m_\mu \equiv t_\mu \Lambda^2 |\Phi_0|^2,
\label{omega_nematic_1}
\end{equation}
which is consistent with Eq.~(\ref{omega_nematic}) if $K_1 = 1 /
m_\tau$, $K_3 = K$, and $\gamma = m_x$.

Remarkably, the magnetophonon mode of the parent
smectic~(\ref{omega_smectic}), does not totally disappear from the
spectrum. Instead, it acquires a small gap $\sqrt{m_y Y}$ at $q = 0$.
This gapped mode anti-crosses with the acoustic
branch~(\ref{omega_nematic_1}) near the point $\omega_1^2({\bf q}) \sim
m_y Y$, and at larger $q$ becomes the lowest frequency collective mode
with the dispersion relation
\begin{equation}
\omega_2({\bf q}) = \left[\frac{q^2_x q^2_y}{m^2 \omega_c^2} Y U(q)
                  + m_y Y\right]^{1/2}
\label{omega_nematic_2}
\end{equation}
only slightly different from~(\ref{omega_smectic}). At such $q$ the
structure factor of the nematic has two sets of $\delta$-functional
peaks,
%
\[
S(\omega, {\bf q}) = \frac{\pi \hbar q_y^2}{m \omega_c^2}
\left[\frac{K q_y^4}{m n_0} \delta(\omega^2 - \omega_1^2) +
\frac{Y q_x^2}{m n_0} \delta(\omega^2 - \omega_2^2)\right],
\]
%
which split between themselves the spectral weight of the single
collective mode of the smectic. The presence of the two modes can be
explained by the existence of two order parameters: the aforementioned
unit vector (more precisely, director) ${\bf N}$ normal to the local
stripe orientation and the complex wavefunction $\Phi_0$ of the
dislocation condensate. Classical 2D nematics have two (overdamped)
modes virtually for the same reason.~\cite{Toner_81}

Recently, Radzihovsky and Dorsey~\cite{Radzihovsky_02} formulated a
qualitatively different theory of the quantum Hall nematics, whose
predictions disagree with our Eqs.~(\ref{L_nematic}) and
(\ref{omega_nematic}). At this point it is unclear whether these authors
study a different kind of nematic or they actually contest the
theoretical models proposed by Balents~\cite{Balents_96} and the present
author.~\cite{Fogler_01} To resolve some of these issues it is
imperative to bring the discussion from the level of effective theory to
the level of quantitative calculations. One promising direction is to
investigate some concrete trial wavefunctions of quantum nematics, e.g.,
Eq.~(\ref{nematic_trial_wavefunction}). Recently, the work in this
direction was continued by Ciftja nad Wexler.~\cite{Wexler_02} It is
also desirable to find a functional form of the electron-electron
interaction which gives rise to the nematic ground state. An educated
guess~\cite{Fogler_xxx} is that even a realistic Coulomb interaction may
be sufficient provided $r_s \ll 1$ and $1 \ll N \ll r_s^{-2}$. However,
the quest for quantum nematic may not be easy. Finite-size study by
Rezayi {\it et al.\/}~\cite{Rezayi_00} suggests that the transition from
the smectic to an isotropic phase as a function of the interaction
parameters (Haldane's pseudopotentials) can also occur via a first-order
transition, without the intermediate nematic phase.

Experimentally, the nematic can be distinguished from the smectic by,
e.g., the microwave absorption technique: the nematic will show two
dispersing collective modes while the smectic will produce a single one.
To circumvent disorder pinning effects, such measurements should be done
at high enough $q$.

\section*{Acknowledgments}

This work is supported by the MIT Pappalardo Fellowships Program in
Physics. I would like to thank A.~A.~Koulakov, B.~I.~Shklovskii, and
V.~M.~Vinokur for previous collaboration on the topics discussed.


\section*{References}
\begin{thebibliography}{29}

\bibitem{FQHE} For review, see {\it Quantum Hall Effect\/} edited by
R.~E.~Prange and S.~M.~Girvin (Springer-Verlag, New York, 1990);
{\it Perspectives in Quantum Hall Effect\/} edited by
S.~Das~Sarma and A.~Pinczuk (Wiley, New York, 1997).

\bibitem{Fogler_xxx} For review, see M.~M.~Fogler,
cond-mat/0111001.

\bibitem{Aleiner_95}I.~L.~Aleiner and L.~I.~Glazman,
\prb {\bf 52}, 11~296 (1995).

\bibitem{Fogler_96} A.~A.~Koulakov, M.~M.~Fogler, and B.~I.~Shklovskii,
\prl {\bf 76}, 499 (1996);
M.~M.~Fogler, A.~A.~Koulakov, and B.~I.~Shklovskii,
\prb {\bf 54}, 1853 (1996).

\bibitem{Fukuyama_79}H.~Fukuyama, P.~M.~Platzman, and P.~W.~Anderson,
\prb {\bf 19}, 5211 (1979).

\bibitem{Fogler_97} M.~M.~Fogler and A.~A.~Koulakov,
\prb {\bf 55}, 9326 (1997).

\bibitem{Moessner_96} R.~Moessner and J.~T.~Chalker,
\prb {\bf 54}, 5006 (1996).

\bibitem{Rezayi_99} E.~H.~Rezayi, F.~D.~M.~Haldane, and K.~Yang,
\prl {\bf 83}, 1219 (1999);
{\it ibid\/} {\bf 85}, 5396 (2000).

\bibitem{Shibata_01} N.~Shibata and D.~Yoshioka,
\prl {\bf 86}, 5755 (2001).

\bibitem{Lilly_99} M.~P.~Lilly, K.~B.~Cooper, J.~P.~Eisenstein,
L.~N.~Pfeiffer, and K.~W.~West,
\prl {\bf 82}, 394 (1999).

\bibitem{Du_99} R.~R.~Du, D.~C.~Tsui, H.~L.~St\"ormer, L.~N.~Pfeiffer,
and K.~W.~West,
Solid\ State\ Commun. {\bf 109}, 389 (1999).

\bibitem{Shayegan_00} M.~Shayegan, H.~C.~Manoharan, S.~J.~Papadakis,
and E.~P.~DePoortere,
Physica\ E\ {\bf 6}, 40 (2000).

\bibitem{Eisenstein_01} J.~P.~Eisenstein, M.~P.~Lilly, K.~B.~Cooper,
L.~N.~Pfeiffer, and K.~W.~West,
Physica\ E\ {\bf 9}, 1 (2001);
J.~P.~Eisenstein,
Solid\ State\ Commun.\ {\bf 117}, 132 (2001).

\bibitem{Fradkin_99} E.~Fradkin and S.~A.~Kivelson,
\prb {\bf 59}, 8065 (1999).

\bibitem{MacDonald_00} A.~H.~MacDonald and M.~P.~A.~Fisher,
\prb {\bf 61}, 5724 (2000).

\bibitem{Fertig_99} H.~A.~Fertig,
\prl {\bf 82}, 3693 (1999).

\bibitem{Fogler_00} M.~M.~Fogler and V.~M.~Vinokur,
\prl {\bf 84}, 5828 (2000).

\bibitem{Cote_00} R.~C\^ot\'e and H.~A.~Fertig,
\prb {\bf 62}, 1993 (2000).

\bibitem{Yi_00} H.~Yi, H.~A.~Fertig, and R.~C\^ot\'e,
\prl {\bf 85}, 4156 (2000).

\bibitem{Fogler_01} M.~M.~Fogler,
cond-mat/0107306.

\bibitem{Wexler_01} C.~Wexler and A.~T.~Dorsey,
\prb {\bf 64}, 115~312 (2001).

\bibitem{Barci_01} D.~G.~Barci, E.~Fradkin, S.~A.~Kivelson, and V.~Oganesyan,
cond-mat/0105448.

\bibitem{Lopatnikova_01} A.~Lopatnikova, B.~I.~Halperin, S.~H.~Simon,
and X.-G.~Wen,
cond-mat/0105079.
 
\bibitem{Musaelian_96} K.~Musaelian and R.~Joynt,
J.\ Phys.\ Cond.\ Mat. {\bf 8}, L105 (1996).

\bibitem{Balents_96} L.~Balents,
Europhys.\ Lett.\ {\bf 33}, 291 (1996).

\bibitem{Wexler_02} O.~Ciftja and C.~Wexler,
cond-mat/0108119.

\bibitem{Radzihovsky_02} L.~Radzihovsky and A.~T.~Dorsey,
cond-mat/0110083.


\bibitem{DeGennes_book} P.~G.~de~Gennes and J.~Prost,
{\it The Physics of Liquid Crystals\/} (Oxford University Press, New York,
1995).

\bibitem{Toner_81} J.~Toner and D.~R.~Nelson,
\prb {\bf 23}, 316 (1981).

\bibitem{Golubovic_92} L.~Golubovi\'c and Z.-G.~Wang,
\prl {\bf 69}, 2535 (1992).

\bibitem{Mazenko_83} G.~F.~Mazenko, S.~Ramaswamy, and J.~Toner,
\prl {\bf 49}, 51 (1982); \pra {\bf 28}, 1618 (1983).

\bibitem{Kats_Lebedev} E.~I.~Kats and V.~V.~Lebedev,
{\it Fluctuational Effects in The Dynamics of Liquid
Crystals\/} (Springer-Verlag, New York, 1994).

\bibitem{Toner_82} J.~Toner,
\prb {\bf 26}, 462 (1982);
A.~R.~Day, T.~C.~Lubensky, and A.~J.~McKane,
\pra {\bf 27}, 1461 (1983).

\bibitem{Fisher_89} M.~P.~A.~Fisher and D.~H.~Lee,
\prb {\bf 39}, 2756 (1989).

\bibitem{Rezayi_00} E.~H.~Rezayi and F.~D.~M.~Haldane,
\prl {\bf 84}, 4685 (2000).

\end{thebibliography}
\end{document}